\documentclass{easychair}
\RequirePackage{latexml}
\RequirePackage[numbers]{natbib}

\RequirePackage{graphicx}
\RequirePackage{url}
\RequirePackage{listings}
\providecommand{\titlerunning}[1]{}
\providecommand{\authorrunning}[1]{}

\begin{document}

\title{WorkingWiki: a MediaWiki-based platform for collaborative research}
\author{Lee Worden\\
McMaster University\\
Hamilton, Ontario, Canada\\
University of California, Berkeley\\
Berkeley, California, United States\\
\url{worden.lee@gmail.com}}

\titlerunning{WorkingWiki}
\authorrunning{Worden}

\maketitle

\begin{abstract} 
WorkingWiki is a software extension for the popular MediaWiki platform
that makes a wiki into a powerful environment for collaborating on
publication-quality manuscripts and software projects.  Developed in
Jonathan Dushoff's theoretical biology lab at McMaster University and
available as free software, it allows wiki users to work together on
anything that can be done by using UNIX commands to transform textual
``source code'' into output.  Researchers can use it to collaborate on
programs written in R, python, C, or any other language, and there are
special features to support easy work on \LaTeX{} documents.  It
develops the potential of the wiki medium to serve as a combination
collaborative text editor, development environment, revision control
system, and publishing platform.  Its potential uses are open-ended
--- its processing is controlled by makefiles that are straightforward
to customize --- and its modular design is intended to allow parts of
it to be adapted to other purposes.
\end{abstract}

Copyright \copyright 2011 by Lee Worden (\url{worden.lee@gmail.com}).  This
work is licensed under the Creative Commons Attribution 3.0
license. The human readable license can be found here:
\url{http://creativecommons.org/licenses/by/3.0}.

\section{Introduction}

The remarkable success of Wikipedia as a collaboratively constructed
repository of human knowledge is strong testimony to the power of the
wiki as a medium for online collaboration.  Wikis --- websites whose
content can be edited by readers --- have been adopted by great
numbers of diverse groups around the world hoping to compile and
present their shared knowledge.

While several somewhat high-profile academic wiki projects have been
launched and later abandoned --- the quantum physics community's Qwiki
\cite{qwiki} for example --- there have also been successful academic
wiki projects that strongly suggest that wikis can be a transformative tool for
accelerating and amplifying the power of research collaborations.  In
particular, the OpenWetWare wiki \cite{oww} has proven itself to have
staying power as a home for an extended research community's projects
and data, and the Polymath project \cite{polymath} provides especially
powerful inspiration regarding the power of online collaboration to
accelerate the process of mathematical discovery.

MediaWiki, the software behind Wikipedia and its sister projects,
available openly as free software, is especially powerful,
full-featured, and stable, and is widely used in academic and
popular sites alike.  It excels at managing information organized into
pages of text, as it is on Wikipedia.  As such, it is very useful for
collaborative and public documentation of a research team's techniques
and results, but not applicable for collaboration on the daily
research itself, which tends to involve writing of software tools for
data analysis and simulation, and production of manuscripts for
publication, with figures, tables, formulae and citations.

This paper describes a software package that extends MediaWiki,
creating a hybrid environment which combines the desirable features of
the wiki system --- easy collaborative editing, recording of history
and authorship, and instant publication on the internet --- with
support for complex formats including programming languages and
\LaTeX{} document formatting, making it possible to collaborate simply
and flexibly on the actual daily work of the lab and making it simple
to store and publish, in an integrated form, the results, process and
presentation of the research.

\section{WorkingWiki}

WorkingWiki \cite{WorkingWiki} is a software extension for the popular
MediaWiki system that makes a wiki into a powerful environment for
collaborating on publication-quality manuscripts and software
projects.  The WorkingWiki extension allows you to store ``source
files" in your wiki and develop, test, run and publish them easily,
along with the products of computations using those source files.
Examples include a project of five \LaTeX{ } files and six EPS images
that compile together into a single PDF file, or an R script that
includes two other R source files and produces a CSV data file and
several EPS figures.  The WorkingWiki extension keeps track of when
the source files have changed and when to redo the processing to
update the output, and how to display the various file formats
involved.  The output files and images can be displayed in wiki pages
along with the source code, and can be used as inputs to further
computations.

\paragraph{Example: collaborating on a \LaTeX{} document.}
It's very common for a group of scientific authors to write a paper by
emailing each other copies of \texttt{.tex} files daily or hourly.
This is inconvenient — in order to look at the paper you have to save
the file into a directory and compile it — and unreliable — it's easy
to get mixed up and lose someone's edits, or overwrite them with
someone else's copy of the file.  One solution is to use a revision
control system such as Subversion \cite{svnbook} or Git \cite{git}
to manage the source code, but if any authors are unwilling to take on
the work of learning to use the tool, they're likely to fall back on
emailing the file or just dictating changes to someone else.
WorkingWiki addresses this problem by providing basic revision control
features together with easy editing.  Once the \texttt{.tex} files,
\texttt{.bib} files, and images are in the wiki, it's easy for
everyone to edit and see the updated results, and the wiki keeps track
of all the changes and their authors, and makes it easy to review or
undo them.  It also provides a convenient place to discuss changes,
without having to put comments into the manuscript itself, and can be
used as a website to present the research to the public.

\paragraph{Example: collaborative, reproducible lab science.}
A research team can use WorkingWiki to archive experimental data
(using the wiki's history features to record who uploaded which data
sets when); develop their data-processing scripts collaboratively in
the wiki; construct the scripts that produce figures and tables in the
wiki; create the manuscript that presents the results in the wiki; and
finally export the manuscript as a \texttt{.tar.gz} file ready to
submit to a scientific journal.  The wiki can then be used to publish
the data, source code, and manuscript to the world as is.  This
process captures all the files needed to understand and reproduce
the research project, with its revision history intact, and in a form
that is easy to annotate and publish online.  A research team
developing simulation programs rather than using experimental data can
use WorkingWiki in the same way.

\vspace{16pt}
\noindent
WorkingWiki is developed principally for research groups, but is
likely to have a variety of other uses as well for mathematicians,
scientists, and software developers.  WorkingWiki provides some
features of an integrated development environment: it coordinates
compiling (if necessary) and running the code when relevant source
files have changed, and displaying the results.  It provides some
features of a revision control system: it uses MediaWiki's history
features to record author, date/time, and content of every change to
the files and the wiki pages they are connected to, and it allows
viewers to export the source code to their workstations and work on it
offline.

It integrates editing and running code with wiki editing.  Source and
product files can be mixed freely into wiki pages' text.  Editing is
fully collaborative.  The effects of changes to source files can be
fully previewed before saving to the wiki.  The WorkingWiki-extended
wiki is a simple, elegant way to present a research team's work to the
world.

WorkingWiki has special features for translating \LaTeX{} documents to
HTML, for display directly in the wiki page.  WorkingWiki allows
collaborators to edit complete \LaTeX{} documents collaboratively on
the wiki, view the compiled document in the wiki page, and export the
documents' source files to your workstation when ready to submit or
circulate.  Using Bruce Miller's \LaTeXML{} software \cite{latexml},
the rendered contents of a \LaTeX{} document are made visible in the
wiki page, including figures, citations and equations (optionally
using MathML), as well as in the standard PDF format.  The editing
history of all files is maintained, including authorship of each
change.  WorkingWiki's \LaTeX{} handling works with documents that
involve multiple files, stored on multiple wiki pages.  \LaTeX{}
\texttt{{\textbackslash}include},
\texttt{{\textbackslash}bibliography},
\texttt{{\textbackslash}includegraphics}, and like commands are
supported.  Filenames do not need to match page names.

WorkingWiki is extensively customizable, supporting collaborative
development and use of computer programs in any language.  Images and
other files created by computer programs can be included directly in
\LaTeX{} documents and read by other programs, and are updated
automatically when the programs or source data files are changed.  The
development environment can be customized by adding default make
rules, and in many other ways.

WorkingWiki supports reproducible and open research by allowing
researchers to collect all the files involved in a research project —
data files, source code, documentation, publications — in an
accessible place where collaborators can develop them together, and
the public can be allowed to download the entire project, to verify
results and try their own experiments.

\section{WorkingWiki in use}

WorkingWiki operates on \emph{source files} that are stored in
standard wiki pages.  Source files are collected into \emph{projects}.
Behind the scenes, WorkingWiki maintains a cached working directory
where it stores and processes the project's files.  When an output
file is called for in a wiki page or by other means, WorkingWiki does
its work by invoking \texttt{make} \cite{make} to create or update the
file from the source files in its project before displaying it.  In
this way, users can edit their code (or their data files, or
\texttt{.tex} documents) by editing the wiki, and run the code and
view the output (the typeset version of the paper, the updated version
of the figure, the textual output of the program) just by previewing
or saving the page.

Figure~\ref{fig:script-example-code} provides a simple example of the
source text of a WorkingWiki-enabled wiki page.  Most of the text of
this page is standard MediaWiki markup using constructs such as
\verb|==|\ldots\verb|==| for section headers and
\verb|[[|\ldots\verb|]]| for links.  A WorkingWiki source file is
defined by including it in an XML-style \texttt{source-file} element.
The text between the opening and closing tags of that element defines
the content of that file, and it is written to the corresponding file
in the project's working directory and used to update the project's
output files as needed.  In this example, project names are not
explicitly given, signaling the software to use by default the project
whose name is the name of the wiki page.  The assignment of files to
projects can be made explicit by supplying a \texttt{project}
attribute along with the \texttt{filename} attribute in the opening
tag.

Below the source file is an output file, represented by a self-closing
\texttt{project-file} element.  When the MediaWiki parser encounters this
tag and passes it to WorkingWiki's code, WorkingWiki synchronizes all
the project's source files with their copies in the working directory,
creates a subprocess to run the Unix command \texttt{make figure.png}, and
(assuming the make command succeeds) retrieves the file and inserts
the file into the HTML page that is the output of the wiki's parser.
Thus simply viewing the page causes the output file to be updated and
displayed.  Figure~\ref{fig:script-example-output} shows what this wiki page
looks like in the web browser.

This example is especially simple because the steps to make the output
from the source code are controlled by a system-wide makefile
installed in a central location and used in all projects.  This is not
necessary: makefiles can also be added to individual projects, in the
same way as any other source file, and in this way users can control
the processing of any kind of files and specify their dependencies.

\begin{figure}
\verb|==R graphics example==|\\
\\
\verb|Here is a simple example of how to do a figure with R, using |\\
\verb![[Recipe_Book#R | the lalashan site's custom rules for using R]].!\\
\verb|The custom rules make it simple: just define a .R file:|\\
\\
\verb|<|\verb|source-file filename=example.R>|\\
\verb|plot(function(x){-x*cos(x-1)}, -pi, pi, col="blue");|\\
\verb|</|\verb|source-file>|\\
\\
\verb|and request its output using just the right filename:|\\
\\
\verb|<|\verb|project-file filename=example.Rout.png/>|

\caption{\label{fig:script-example-code}Source text for example
WorkingWiki-enabled wiki page, illustrating the use of
the \texttt{source-file} and \texttt{project-file} tags.}
\end{figure}

\begin{figure}
\begin{center}
\includegraphics[width=0.9\textwidth]{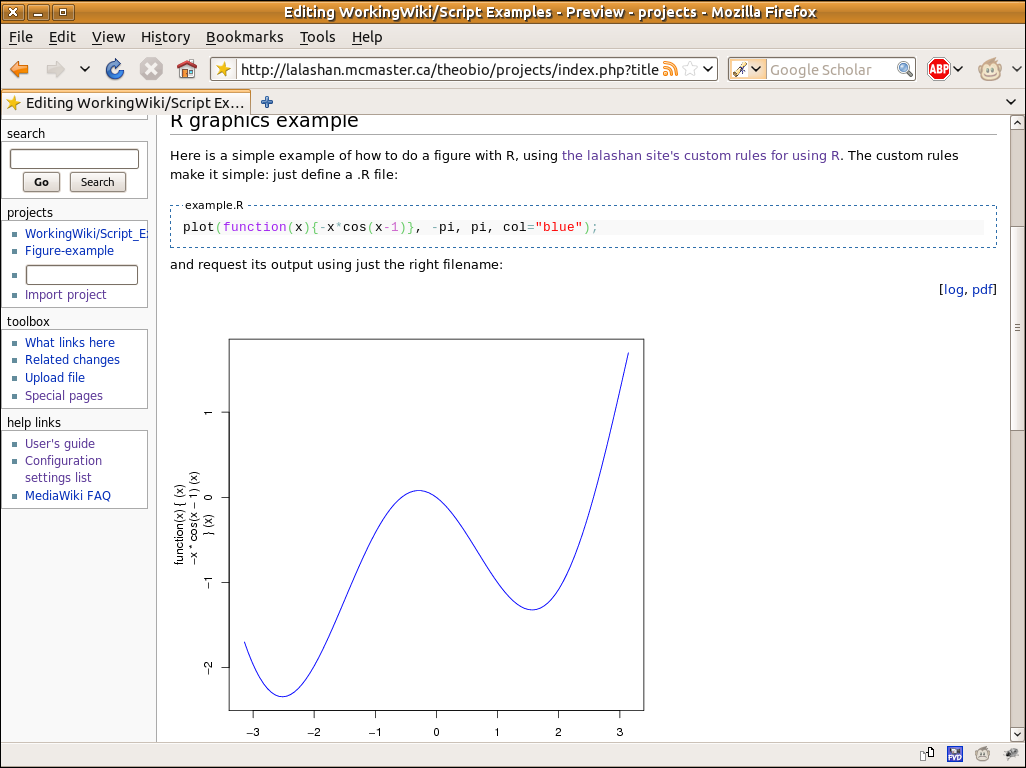}
\end{center}
\caption{\label{fig:script-example-output}
How the markup of figure~\ref{fig:script-example-code} appears in a wiki
page.}
\end{figure}

WorkingWiki completely supports MediaWiki's previewing feature:
changes to pages, including source files, can be tested and revised
extensively before saving them to the wiki.  When a user previews a
page that includes project files, WorkingWiki updates them from the
modified source files, in a separate preview copy of the project's working
directory.  When the user saves the changes to the wiki, WorkingWiki
merges the temporary files into the permanent copy, to avoid
unnecessary repetition of processing steps. \footnote{When a directory
becomes large, these copy operations can become quite expensive.
Unfortunately, it's necessary, because if we processed unsaved code in
the primary cache directory it could modify files in ways that would
affect the outcome of future processing steps, even if the previewed
changes were never saved.  We partially address this problem by making
it possible to split out large or numerous project files into separate
project directories that are left uncopied provided the project's
authors promise that they are protected from problematic side effects,
but a more flexible solution would be desirable.  We are considering
using the Btrfs filesystem's copy-on-write file storage capabilities
\cite{btrfs} to make these copy and merge operations fast and cheap.}

Editing source code in a form field in a web browser is much less
convenient than editing files in full-featured editors like vi and
emacs, but browser addons It's All Text \cite{itsalltext} for Firefox
and TextAid \cite{textaid} for Chrome make it much easier by allowing
a page's contents to be opened in a text editor of the user's choice
and kept open while repeatedly submitting and previewing the page or a
section of the page.  This gives users access to all the editor
features they are used to, such as syntax highlighting and smart
indenting.

\subsection{\LaTeX{} features}

WorkingWiki allows a source file to be displayed in a transformed
form.  For instance, if a page's editor writes
\verb|<|\verb|source-file filename="example.R" display="example.Rout.png">|, the project
file \texttt{example.Rout.png} is updated and displayed in the page in
place of the source code.  This feature has not proved very popular
--- it seems to be preferable to make source code visible in most cases
--- but the use of \emph{default} display attributes is very useful
with \LaTeX{} and related formats.  Default display transformations
can be defined by the wiki's administrators, and WorkingWiki comes
with a small number of them predefined.  In particular, \texttt{.tex}
files are by default transformed to \texttt{.latexml.html} files for
display, and WorkingWiki's system-wide makefile provides rules that
make that transformation by using \LaTeXML{} to process the document
into HTML for display.  Additionally, if a user has opted to enable
MathML output and is using a MathML-compatible browser, WorkingWiki
instead provides a \texttt{.latexml.xhtml} version of the document
which uses MathML for all mathematical content.  (This automatic
detection and output switching is also available to wiki users and
administrators for custom HTML-producing processes.)

When displaying a \texttt{.tex} file, WorkingWiki also provides a link
that makes and provides a PDF version of the document using either
\LaTeX{} or PDF\LaTeX{} (or other programs, if customized).  This link can
be redefined or additional links can be added to the default behaviour
and they can be changed on a file-by-file basis by adding attributes
to the \texttt{source-file} element.

\begin{figure}
\begin{center}
\includegraphics[width=0.9\textwidth]{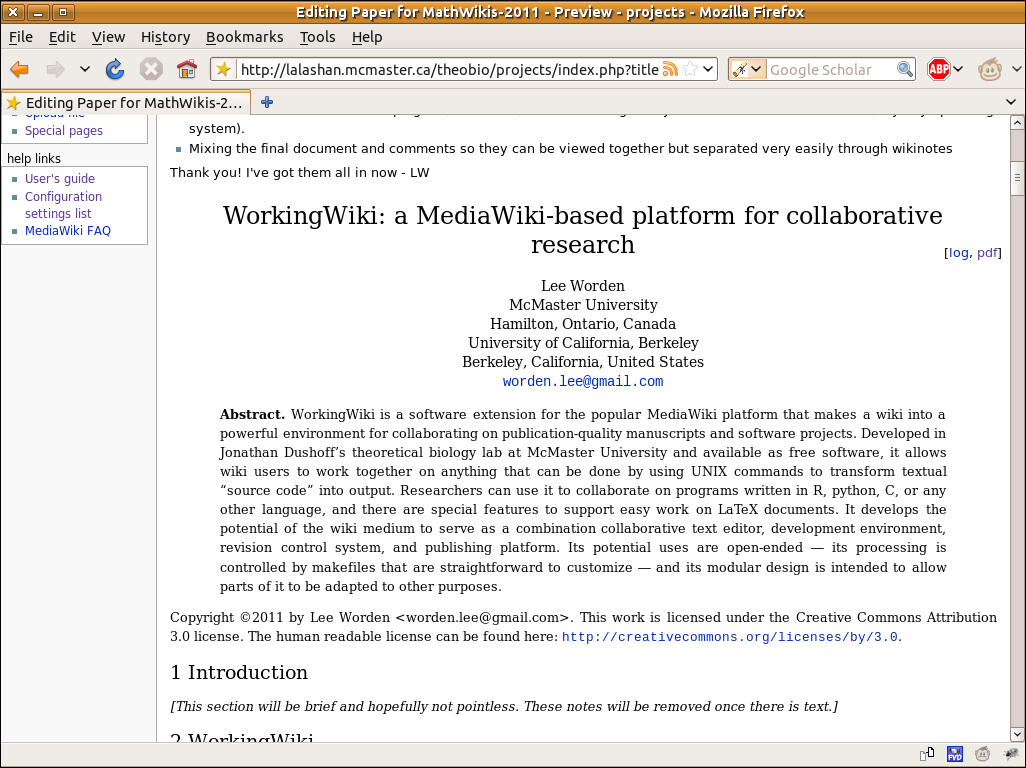}
\end{center}
\caption{\label{fig:latex-example-output}
A \LaTeX{} document in a wiki page.}
\end{figure}

In figure~\ref{fig:latex-example-output} is a screenshot of a \LaTeX{}
document in a wiki page, illustrating how the XHTML version of the
manuscript is embedded in the page, and the PDF link at the right
margin.  Embedding the rendered form of the manuscript directly in the
wiki page allows for a comfortable cycle of editing, previewing, and
editing some more, which is comparable to the ease of editing a
manuscript's source text in a text editor, processing it into DVI or
PDF, viewing, and editing again, especially when using an external
text editor with the wiki as described above.

These features allow users to edit \LaTeX{} documents (and other
source files) in the same way that wiki pages are edited: open an edit
form and change the source code; press the preview button and see what
it looks like when processed; edit some more until it is right, and
then save.

The make rules that are provided with WorkingWiki automatically keep
track of dependencies on BibTeX files, figures, locally provided style
files and included tex files, so that the displayed manuscript is kept
up to date when any part of it is updated.

It is simple to use \LaTeX{} conditionals to insert comments and
conversations in the code of a manuscript that are visible in the HTML
version of the manuscript but invisible in the PDF, providing an easy
way to coordinate while keeping a clean manuscript for submission.

\subsection{Advanced features}

\subsubsection{Inter-project dependencies}

For advanced users, WorkingWiki supports sharing of data among multiple
projects, and takes steps to ensure dependency relationships are
respected and data integrity is protected when previewing or running
background jobs (see below).

This feature allows a number of useful strategies.  General-use code
can be shared among multiple projects, by placing it in a "library"
project.  Complex projects can be organized by grouping related things
together into separate WorkingWiki projects, while allowing
interaction between the different components.  Independent parts of a
project can be isolated from one another.  A particularly important
case is that a journal article for publication can be housed in a
separate project from the data and programs that provide its content.
This allows, on the one hand, the authors to maintain the dependency
relationships within the wiki that allow the manuscript's figures and
tables to be automatically kept up to date when the data and programs
change, and on the other hand makes it simple to export the article's
source files in a neat \texttt{.tar.gz} package for submission to the
journal and leave the programs and data behind.

\subsubsection{Interaction with external data}

WorkingWiki's back end is capable of processing data from multiple
sources, and the front end allows those projects to be integrated with
projects originating on the wiki.  For instance, this author has a
research project in progress in which a complex simulation program is
stored in two GitHub repositories, pulled into two project directories in the wiki's file
cache, compiled and run by code in a third project whose source files
are housed on the wiki, and the output files are stored in a fourth
project that is created on the wiki but doesn't have any source files.

This external-project feature also allows interaction between projects
housed on different wikis --- this is useful on our site at McMaster
because we operate many interconnected wikis, and store some
general-use code on a central wiki for use on others.

Project data can be exported to a user's local disk in a
\texttt{.tar.gz} package, which includes the wiki's centralized
makefile and other supplementary data, to allow running and
developing the code offline.  It can then be re-imported into the
wiki.  There is also a command-line tool to pull project files from
wikis to a local directory.  In the future there may be an interface
to git \cite{git}, allowing one to pull and push source code from and
to the wiki storage as if it were a (somewhat simplified) git
repository.  Given the flexibility of the git client, this would
effectively make it possible to migrate projects easily between wiki
storage and many other repositories.

\subsubsection{Background jobs}

When certain computational steps are too slow to run on the spot,
WorkingWiki allows them to be run as background jobs, which run
outside of the wikitext-parsing process.  A background job is created
simply by specifying a make target and requesting it be made in the
background.  Any background jobs that have been created are listed at
the top of all pages that interact with the projects they involve, in
a listing that provides their basic information and status.  Whether a
job has succeeded or failed, a user can browse its files, destroy it,
or merge its output into the project's primary working directory.
Running jobs can be browsed and killed.

In a standard installation, background jobs are run as Unix
subprocesses on the same processors as the web server (using
\texttt{nice} and \texttt{ionice} at the discretion of the site
administrators), but there is a prototype in development to run
background jobs on computing clusters using GridEngine \cite{sge}.

\section{Design of the software}

WorkingWiki is implemented as a MediaWiki extension, written in PHP
and augmented by a few JavaScript and CSS resources, makefiles, and
small helper programs.  It is freely available under the GNU General
Public License \cite{gpl2}, and is compatible with all versions of
MediaWiki from 1.13 on.

The \texttt{source-file} and \texttt{project-file} tags are
implemented as tag hooks, a standard means of extending MediaWiki's
parser, and retrieval of binary files and project management are
provided by two special pages, another standard form for extensions.
Like MediaWiki, WorkingWiki itself provides a number of hooks that can
be used by other extensions to provide additional features or modify
the ones that are provided.  It has not been tested in combination
with all other MediaWiki extensions, but there are no known conflicts.

WorkingWiki's behavior can be extensively customized as is.
Administrators can modify the rules controlling how different file
types are displayed, and provide default transformations like the one
from \LaTeX{} to HTML and links like the one from \LaTeX{} to PDF.
Custom make rules can be added, to make it easy for users to write
source code and transform it in standard ways, and the existing make
rules can be partially or completely overridden.

\subsection{Separation of wiki from project engine} 
A wiki is a powerful tool that combines a number of important
functions.  It is effectively a combined revision control system,
integrated development environment, markup parser for website content,
and publishing platform for web pages written in its markup language.
WorkingWiki extends all of these functions to a wider range of source
material, making the wiki into a combination revision control system,
development environment, execution environment, and publishing
platform for the general case of executable program text.  Each of
these functions is provided in more powerful forms by other tools, but
the power of the wiki medium is in combining them together in an
elegant, easy-to-use form.

An ideal situation would be to make it easy for end users to separate
all these functions in a mix-and-match way, for instance providing a
development, execution and publishing platform for data stored in a
revision control system of the user's choice, or providing revision
control, execution and publishing but using a third-party tool for
editing and previewing.  This is not entirely possible at present, but
WorkingWiki is written with these separations in mind.

In particular, while the revision control, development (e.g. editing
and previewing), and publishing functions are essentially provided by
MediaWiki once the source files' contents are inserted into the stored
pages and output files are inserted into the output HTML,
WorkingWiki's execution environment is entirely separate from
MediaWiki's code, and is designed as a completely independent
component.

This component, called ProjectEngine, is a standalone tool that stores
files, performs make operations, and serves up-to-date file contents.
Written in PHP, as are MediaWiki and WorkingWiki, it can be used as a
component of a larger program --- it is incorporated in WorkingWiki in
this way by default --- and can also run as a self-contained HTTP
service.  It can be thought of as similar to a simple web server ---
whose primary function is to retrieve the contents of files for
clients --- but one that can create and update its files using make
rules before serving them.

ProjectEngine supports updating and removing files; creating,
destroying and merging preview sessions by making a copy of
"persistent" files; and creating, destroying, merging and tracking
background jobs.

The project engine seems to be a simple and powerful concept, and one
that may have uses beyond this single wiki system.  If nothing else,
it can be used as a back end for similar extensions for other wiki
engines, and the author has discussed this possibility with the author
of Projects Wiki \cite{projectswiki}, a WorkingWiki-inspired plugin
for Dokuwiki.

\subsection{Security considerations} 

There are, of course, risks involved in running a web server that
includes a project engine, which executes programs supplied by users.
To a first approximation, the risks can be partitioned into just a few
categories: overuse or destruction of server resources, access to
sensitive data, denial of service, and harmful output.  All of these
risks can be managed.

The first category, overuse or destruction of system resources, is
fairly broad.  It includes scenarios from user-supplied code altering
files on the server, to programs that send voluminous spam emails to
innocent people, to infinite loops that consume excessive CPU time or
fill up a disk partition.  These risks can be managed by use of a
mandatory access control system such as TOMOYO Linux \cite{tomoyo} to
restrict access to all system resources, from sensitive files to use
of the server's network interfaces.  Additionally, ProjectEngine uses
\texttt{nice} and \texttt{ionice} to prevent its processes from
monopolizing CPU time and disk access, and uses \texttt{setrlimit()}
to limit the number of subprocesses a make process can create and
kills make processes after a limited time period.  A quota system can
be used to limit the amount of disk space ProjectEngine's files can
consume.

Mandatory access control is also effective at keeping user-supplied
code from reading sensitive system files, and WorkingWiki's inputs are
carefully validated to prevent backdoor access to SQL data.
WorkingWiki works with MediaWiki's access control features to ensure
that a password-protected wiki doesn't reveal data to unauthorized
users.

Denial of service attacks can include inputs that cause crashes in the
software, as well as inputs that consume inordinate resources.  The
latter category has been covered.  The former case can probably never
be ruled out with complete confidence, but in any case, when a wiki is
password-protected, unauthorized users have no means to interact with
WorkingWiki and thus any attacks from outside the user community must
be directed at other services.

The case of harmful output is the least well accounted for at present:
in order to provide an HTML rendering of \LaTeX{} documents, it's
necessary to allow ProjectEngine jobs to produce HTML output to be
passed on to the client, and in order to support programmers it's
necessary to allow them to write programs and custom make rules; in
combination this means that users' projects can produce HTML output
that does unwelcome things on the client side, such as making calls to
third-party websites that reveal information about users logged in to
the wiki.  It may be possible to filter HTML output in a way that
allows only safe output, but this is currently not implemented in
WorkingWiki.  Another possibility is to provide as an option a
restricted set of WorkingWiki features, for instance allowing users to
edit \LaTeX{} documents but not to create makefiles; this might
suffice to provide a system that could be safely opened up to
anonymous editors.

The current recommendation is to use WorkingWiki only on
password-protected wikis, restricting editing access to trusted users.
We believe it is safe for publicly readable wikis as long as only
trusted users can edit.  WorkingWiki is very useful and reliable for
semi-closed wikis in this way, and use in public wikis more like
Wikipedia may be possible in the future.

\section{Examples of WorkingWiki in use}

WorkingWiki's home site \cite{WorkingWiki}, which is itself a
WorkingWiki-enabled wiki, provides a handful of example WorkingWiki
projects, illustrating how to create projects for \LaTeX{} and for
programming (the nomogram example \cite{ww-nomogram} is especially
engaging).

One active research team is using it to analyze and visualize African
survey data related to HIV and female genital cutting. For that
research a utility project has been created that automates the process
of downloading the raw survey data from the provider's web site,
merging separate data sets together, and transforming them into
\texttt{.RData} files ready for processing in R.  Another utility
project provides custom R functions for plotting the data, allowing
users to create visualizations of particular variables, including
geographical plots, by inserting brief scripts of only a few lines
into their wiki pages.

The Dushoff lab has created a suite of make rules to streamline R
programming within WorkingWiki, making it easy to process data in
steps by creating a series of small R modules that operate on the data
produced by earlier modules, and interleaving these brief program
snippets with the plots and textual output that they produce, in a
wiki page that documents the steps of the processing.  This mode of
working with processing steps and their output is similar to the
interactive notebooks provided by Mathematica \cite{Mathematica} or
Sage \cite{sagemath}.  The Dushoff lab has also developed custom
processes and make rules for automatically generating BibTeX data and
browser-friendly reference lists from PubMed and similar identifiers,
making it easy to maintain citation data within the wikis.

This author is conducting an experiment in open research by
maintaining a project on a publicly readable wiki.  This project,
which is in the early proof-of-concept phase, combines simulation and
mathematical analysis in modeling collective search for a solution to
a complex problem \cite{consensus}.

Another team is using it to investigate the behavior of spatially
extended threshold models like those described by Schelling
\cite{Schelling} and Granovetter \cite{Granovetter}, using a
combination of python simulations and collaborative mathematical
analysis (both in WorkingWiki).  Other teams are using WorkingWiki to
study the use of non-negative matrix factorization for community
detection in marine ecometagenomics data, the effect of complex
contact network structure on infectious disease dynamics, and spread
of coexisting favorable mutations in spatially localized populations
of plants and animals.

Papers completed in WorkingWiki have been published in
\emph{Ecological Economics} \cite{Greenhouse}, \emph{Theoretical
  Ecology} \cite{He+}, and \emph{Journal of Mathematical Biology}
\cite{Jiang+} (and now in this proceedings \cite{this}).

\section{WorkingWiki and math wikis}

WorkingWiki's makefile rules are straightforward to extend or replace.
Its \LaTeX{} features can be extended to additional formats: this
author once created a structure for working on Sweave documents in
WorkingWiki in an afternoon (it took a few minutes to write the rule
to create \texttt{.tex} files automatically from Sweave files, and a
few hours to create a \LaTeXML{} style file to make the Sweave output
look good in the browser).  It should be straightforward to extend WorkingWiki to
process any number of specialized document types conveniently; for
instance, allowing users to edit sTex \cite{stex} documents and render
them automatically to PDF, browser-ready XHTML, and OMDoc \cite{omdoc}
formats.  It could also be used to develop, store, and process
documents in computer-aided theorem-proving systems, or using any
other tool that can be invoked from a UNIX command line to process
files.  Its uses are open-ended, and may prove very fruitful to
explore.

\bibliographystyle{plain}
\bibliography{worden-mwitp}

\end{document}